\documentclass[12pt,nohyper,notoc]{JHEP}
 
\usepackage{amssymb} 
\usepackage{latexsym}

\newcounter{multieqs}

\newcommand{\bq}{\begin{equation}}
\newcommand{\fq}{\end{equation}}
\newcommand{\bqr}{\begin{eqnarray}}
\newcommand{\fqr}{\end{eqnarray}}

\newcommand{\be}{\begin{equation}}
\newcommand{\ee}{\end{equation}}
\newcommand{\eq}[1]{(\ref{#1})}

\newcommand{\bm}[1]{\mbox{\boldmath $#1$}}

\def\bd{\begin{document}}
\def\ed{\end{document}}
\def\nn{\nonumber}
\def\bea{\begin{eqnarray}}
\def\eea{\end{eqnarray}}
\let\bm=\bibitem
\let\la=\label

\def\reps{representations }

\def\npb#1#2#3{Nucl. Phys. {\bf{B#1}} #3 (#2)}
\def\plb#1#2#3{Phys. Lett. {\bf{#1B}} #3 (#2)}
\def\prl#1#2#3{Phys. Rev. Lett. {\bf{#1}} #3 (#2)}
\def\prd#1#2#3{Phys. Rev. {D \bf{#1}} #3 (#2)}
\def\cmp#1#2#3{Comm. Math. Phys. {\bf{#1}} #3 (#2)}
\def\cqg#1#2#3{Class. Quantum Grav. {\bf{#1}} #3 (#2)}
\def\nppsa#1#2#3{Nucl. Phys. B (Proc. Suppl.) {\bf{#1A}}#3 (#2)}
\def\ap#1#2#3{Ann. of Phys. {\bf{#1}} #3 (#2)}
\def\ijmp#1#2#3{Int. J. Mod. Phys. {\bf{A#1}} #3 (#2)}
\def\rmp#1#2#3{Rev. Mod. Phys. {\bf{#1}} #3 (#2)}
\def\mpla#1#2#3{Mod. Phys. Lett. {\bf A#1} #3 (#2)}
\def\jhep#1#2#3{J. High Energy Phys. {\bf #1} #3 (#2)}
\def\atmp#1#2#3{Adv. Theor. Math. Phys. {\bf #1} #3 (#2)}

\newcommand{\EQ}[1]{\begin{equation} #1 \end{equation}}
\newcommand{\AL}[1]{\begin{subequations}\begin{align} #1 \end{align}\end{subequations}}
\newcommand{\SP}[1]{\begin{equation}\begin{split} #1 \end{split}\end{equation}}
\newcommand{\ALAT}[2]{\begin{subequations}\begin{alignat}{#1} #2 \end{alignat}\end{subequations}}
\def\beqa{\begin{eqnarray}} 
\def\eeqa{\end{eqnarray}} 
\def\beq{\begin{equation}} 
\def\eeq{\end{equation}} 

\def\N{{\cal N}}
\def\sst{\scriptscriptstyle}
\def\thetabar{\bar\theta}
\def\Tr{{\rm Tr}}
\def\one{\mbox{1 \kern-.59em {\rm l}}}

\def\a{\alpha}          \def\da{{\dot\alpha}}
\def\b{\beta}           \def\db{{\dot\beta}}
\def\c{\gamma}  \def\C{\Gamma}  \def\cdt{\dot\gamma}
\def\d{\delta}  \def\D{\Delta}  \def\ddt{\dot\delta}
\def\e{\epsilon}                \def\vare{\varepsilon}
\def\f{\phi}    \def\F{\Phi}    \def\vvf{\f}
\def\h{\eta}
\def\k{\kappa}
\def\l{\lambda} \def\L{\Lambda}
\def\m{\mu}     \def\n{\nu}
\def\p{\pi}     \def\P{\Pi}
\def\r{\rho}
\def\s{\sigma}  \def\S{\Sigma}
\def\t{\tau}
\def\th{\theta} \def\Th{\Theta} \def\vth{\vartheta}
\def\X{\Xeta}
\def\z{\zeta}

\def\cA{{\cal A}} \def\cB{{\cal B}} \def\cC{{\cal C}}
\def\cD{{\cal D}} \def\cE{{\cal E}} \def\cF{{\cal F}}
\def\cG{{\cal G}} \def\cH{{\cal H}} \def\cI{{\cal I}}
\def\cJ{{\cal J}} \def\cK{{\cal K}} \def\cL{{\cal L}}
\def\cM{{\cal M}} \def\cN{{\cal N}} \def\cO{{\cal O}}
\def\cP{{\cal P}} \def\cQ{{\cal Q}} \def\cR{{\cal R}}
\def\cS{{\cal S}} \def\cT{{\cal T}} \def\cU{{\cal U}}
\def\cV{{\cal V}} \def\cW{{\cal W}} \def\cX{{\cal X}}
\def\cY{{\cal Y}} \def\cZ{{\cal Z}}

\def\ua{\underline{\alpha}}
\def\ub{\underline{\phantom{\alpha}}\!\!\!\beta}
\def\uc{\underline{\phantom{\alpha}}\!\!\!\gamma}
\def\um{\underline{\mu}}
\def\ud{\underline\delta}
\def\ue{\underline\epsilon}
\def\una{\underline a}\def\unA{\underline A}
\def\unb{\underline b}\def\unB{\underline B}
\def\unc{\underline c}\def\unC{\underline C}
\def\und{\underline d}\def\unD{\underline D}
\def\une{\underline e}\def\unE{\underline E}
\def\unf{\underline{\phantom{e}}\!\!\!\! f}\def\unF{\underline F}
\def\unm{\underline m}\def\unM{\underline M}
\def\unn{\underline n}\def\unN{\underline N}
\def\unp{\underline{\phantom{a}}\!\!\! p}\def\unP{\underline P}
\def\unq{\underline{\phantom{a}}\!\!\! q}
\def\unQ{\underline{\phantom{A}}\!\!\!\! Q}
\def\unH{\underline{H}}

\def\As {{A \hspace{-6.4pt} \slash}\;}
\def\Ds {{D \hspace{-6.4pt} \slash}\;}
\def\ds {{\del \hspace{-6.4pt} \slash}\;}
\def\ss {{\s \hspace{-6.4pt} \slash}\;}
\def\ks {{ k \hspace{-6.4pt} \slash}\;}
\def\ps {{p \hspace{-6.4pt} \slash}\;}
\def\pas {{{p_1} \hspace{-6.4pt} \slash}\;}
\def\pbs {{{p_2} \hspace{-6.4pt} \slash}\;}

\def\Fh{\hat{F}}
\def\Xh{\hat{X}}
\def\ah{\hat{a}}
\def\xh{\hat{x}}
\def\yh{\hat{y}}
\def\ph{\hat{p}}
\def\xih{\hat{\xi}}

\def\psit{\tilde{\psi}}
\def\Psit{\tilde{\Psi}}
\def\tht{\tilde{\th}}
 
\def\At{\tilde{A}}
\def\Qt{\tilde{Q}}
\def\Rt{\tilde{R}}

\def\ft{\tilde{f}}
\def\pt{\tilde{p}}
\def\qt{\tilde{q}}
\def\vt{\tilde{v}}

\def\delb{\bar{\partial}}
\def\bz{\bar{z}}
\def\Db{\bar{D}}

\def\bR{{\bf{R}}}

\def\d{\delta}\def\D{\Delta}\def\ddt{\dot\delta}

\def\pa{\partial} \def\del{\partial}
\def\xx{\times}

\def\trp{^{\top}}
\def\inv{^{-1}}
\def\dag{^{\dagger}}
\def\pr{^{\prime}}

\def\rar{\rightarrow}
\def\lar{\leftarrow}
\def\lrar{\leftrightarrow}

\newcommand{\0}{\,\!}      
\def\one{1\!\!1\,\,}
\def\im{\imath}
\def\jm{\jmath}

\newcommand{\tr}{\mbox{tr}}
\newcommand{\slsh}[1]{/ \!\!\!\! #1}

\def\vac{|0\rangle}
\def\lvac{\langle 0|}

\def\hlf{\frac{1}{2}}
\def\ove#1{\frac{1}{#1}}
\def \hf{\frac 12} 

\def\Box{\square}
\def\ZZ{\mathbb{Z}}
\def\CC#1{({\bf #1})}
\def\bcomment#1{}
\def\bfhat#1{{\bf \hat{#1}}}
\def\VEV#1{\left\langle #1\right\rangle}

\newcommand{\ex}[1]{{\rm e}^{#1}} \def\ii{{\rm i}}

\title{Scaling Limits of the Fuzzy Sphere at one Loop}

\author{Chong-Sun Chu$^{a}$, John Madore$^{b,c}$ 
and Harold Steinacker$^{b,d}$\\
${}^a$Centre for Particle Theory, Department of Mathematical Sciences,\\
University of Durham, Durham, DH1 3LE, UK\\[2ex]
${}^b$ Laboratoire de Physique Th\'eorique et Hautes Energies\\
        Universit\'e de Paris-Sud, B\^atiment 211, F-91405 Orsay \\[2ex]
${}^c$ Max-Planck-Institut f\"ur Physik\\
           F\"ohringer Ring 6, D-80805 M\"unchen  \\[2ex]
${}^d$ Sektion Physik der Ludwig--Maximilians--Universit\"at M\"unchen\\
        Theresienstr.\ 37, D-80333 M\"unchen  \\[1ex]
E-mail: {\tt chong-sun.chu@durham.ac.uk,
Harold.Steinacker@th.u-psud.fr, John.Madore@th.u-psud.fr}
}

\abstract{We study the one loop dynamics of QFT on the fuzzy sphere
and calculate the planar and nonplanar contributions to
the two point function at one loop.
We show that there is no UV/IR mixing on the fuzzy sphere.
The fuzzy sphere is characterized by two moduli: a 
dimensionless parameter $N$ and a dimensionful radius $R$. 
Different geometrical phases can obtained at 
different corners of the moduli space. 
In the limit of the commutative sphere, we find 
that the two point function is regular without UV/IR mixing; however 
quantization does not commute with the commutative limit,
and a finite ``noncommutative anomaly'' survives in the commutative limit.
In a different limit, the noncommutative plane $\bR^2_\th$ is
obtained, and the UV/IR mixing reappears. This 
provides an explanation of the UV/IR mixing 
as an infinite variant of the ``noncommutative anomaly''.}

\keywords{Non-Commutative Geometry, Quantum Effective Action}

\preprint{{\tt 
               LMU-TPW 07/01}\\
            { \tt  LPT-ORSAY 01-61 }}

\begin{document}
\section{Introduction}

Much effort has been spent in recent years to study quantum field 
theory on noncommutative spaces. There are many reasons why this is of 
interest, most of which are related to our 
poor understanding of physics and the nature of spacetime 
at very short distances. Additional motivation came from the
possibility to realize such spaces in
string theory or M-theory \cite{bfield,cds,alekseev,myers}.
This provides new insights to issues such as
nonlocality and causality of a noncommutative field theory, 
which are crucial in
understanding the structure of quantum spacetime.  

Some of the problems that can arise in QFT on noncommutative spaces are 
illustrated in the much--studied example of the noncommutative plane
$\bR^n_\th$; see \cite{dn} for a recent
review. One of the most intriguing phenomena on that space
is the existence of an ultraviolet/infrared 
(UV/IR) mixing \cite{uvir1} in the quantum effective 
action. Due to this
mixing, an IR singularity arises from integrating out the UV degrees of
freedom. This threatens the renormalizability and even the
existence of a QFT. 
Hence a better
understanding (beyond the technical level) of the mechanism of UV/IR mixing
and possible ways to resolve it
are certainly highly desirable. 
One possible approach is to approximate
$\bR^n_\th$ in terms of a different noncommutative space.
We realize this idea in the present paper, 
approximating $\bR^2_\th$ by a fuzzy sphere.
This will allow to understand the UV/IR mixing as an infinite variant of a
``noncommutative anomaly'' on the fuzzy sphere,
which is a closely related but different phenomenon
discussed below. This is one of our main
results. 
A related, but less geometric approach was considered in \cite{kinar}.

In this article, we consider scalar $\Phi^4$ theory on the fuzzy sphere,
and  calculate the two point function at one loop.
The fuzzy sphere $S^2_N$
is a particularly simple noncommutative space \cite{madore},
characterized by its radius $R$ and a ``noncommutativity'' parameter
$N$ which is an integer. It  approaches 
the classical sphere in the limit $N \rightarrow \infty$ for fixed $R$,
and can be thought of as consisting of $N$ ``quantum cells''.
The algebra of functions on $S^2_N$ is finite, with maximal angular 
momentum $N$. Nevertheless, it admits the full symmetry group $SO(3)$ 
of motions. The fuzzy sphere is closely related to several other
noncommutative spaces \cite{h1}. In particular, it can be 
used as an approximation to the quantum plane $\bR^2_\th$, by
``blowing up'' for example the neighborhood of the south pole. Thus QFT
on $S^2_N$ should provide an approximation of the QFT on the
quantum plane. 

The fuzzy sphere has the additional merit that it is very clear how to 
quantize field theory on it, using a finite analog of the path
integral \cite{gkp}. Therefore QFT on this space is a priori 
completely well--defined, on a mathematical level. 
Nevertheless, it is not clear
at all whether such a theory makes sense from a
physical point of view, i.e. whether there exists a limiting theory 
for large $N$, which could be interpreted as a 
QFT on the classical sphere. There might be a similar UV/IR 
problem as on the quantum plane $\bR^2_\th$, as was claimed
in a recent paper \cite{vaidya}. In other words, it is not clear if 
and in what sense such a QFT is renormalizable. 
As a first step, we calculate in the present paper
the two point function at one loop and find that it is well
defined and regular, without UV/IR mixing. 
Moreover, we find a closed formula for 
the two point function in the commutative limit, i.e. we calculate the
leading term in a $1/N$ expansion. 

It turns out that the 1--loop effective action on $S^2_N$
in the commutative limit differs 
from the 1--loop effective action on the commutative sphere $S^2$
by a finite term, which we call  ``noncommutative anomaly'' 
(NCA). It is a mildly nonlocal, ``small'' correction to the kinetic energy 
on $S^2$, and changes the dispersion relation. 
It arises from the nonplanar loop integration.
Finally, we consider the planar limit of the fuzzy sphere. 
We find that a IR singularity is
developed in the nonplanar two point function, and hence the UV/IR mixing
emerges in this limit. 
This provides an understanding of the UV/IR mixing for QFT on $\bR^2_\th$
as a ``noncommutative anomaly'' which 
becomes singular in the planar limit of the fuzzy sphere dynamics.

This paper is organized as follows.
In section 2, we consider different geometrical limits of
the classical (ie. $\hbar =0$) fuzzy sphere. In particular, we show how the
commutative sphere and the noncommutative plane $\bR^2_\th$
can be obtained  in different corners of the moduli space of the fuzzy sphere.
In section 3, we study the quantum 
effects of scalar $\Phi^4$ field theory on the fuzzy sphere at 1-loop. 
We show that the planar and nonplanar 2-point function are both regular in the
external angular momentum and no IR singularity is developed.
This means that there is no UV/IR mixing phenomenon on the fuzzy sphere.
We also find that the planar and nonplanar two point functions differ 
by a finite amount which is smooth in the external angular momentum,
and survives in the commutative limit. Therefore the commutative
limit of the $\Phi^4$ theory at one loop differs from the
corresponding one loop quantum theory on the commutative sphere 
by a finite term \eq{NCA}.
In section 4, we consider the planar limit of this QFT, 
and recover the UV/IR mixing.

\section{The Fuzzy Sphere and some Limits}

\subsection{The fuzzy sphere $S^2_N$}

We start by recalling the definition of fuzzy sphere in order to fix
our conventions and notation. The algebra of
functions on the fuzzy sphere is the finite algebra $S_N^2$
generated by Hermitian operators 
$x= (x_1, x_2, x_3)$ satisfying the defining relations
\bea 
[x_i, x_j] = i \l_N \e_{ijk} x_k, \label{def1}\\
x_1^2 + x_2^2 +x_3^2  = R^2. \label{def2}
\eea
The noncommutativity parameter $\l_N$ is of dimension length, and
can be taken positive. The radius $R$ is quantized in units of $\l_N$ by
\be\label{def3}
\frac {R}{\l_N} = \sqrt{\frac{N}{2} \left( \frac{N}{2} +1\right)
} \; , 
\quad \mbox{$N = 1,2,\cdots$ } 
\ee
This quantization can be easily understood. Indeed \eq{def1} is
simply the Lie algebra $su(2)$, whose irreducible representation
are labeled by the spin $\a:= N/2$. The Casimir 
of the spin-$N/2$ representation is quantized, and related to $R^2$ by 
\eq{def2}.
Thus the fuzzy sphere is characterized by its radius $R$ and the
``noncommutativity parameters'' $N$ or $\l_N$. 
The algebra of ``functions'' $S_N^2$ is simply the
algebra $Mat(N+1)$ of $(N+1) \times (N+1)$ matrices. 
It is covariant under the adjoint action of $SU(2)$, under which it 
decomposes into the irreducible \reps with dimensions
$(1) \oplus (3) \oplus (5) \oplus ... \oplus (2N+1)$.

The integral of a function $F \in S_N^2$ over the fuzzy sphere is given by
\be
R^2 \int F  = \frac{4 \pi R^2}{N+1} \tr[ F(x)],
\ee
where we have introduced $\int$, the integral over the fuzzy sphere with 
unit radius.
It agrees with the integral $\int d \Omega$  on $S^2$ in the large $N$ limit. 
Invariance of the integral under  the rotations 
$SU(2)$ amounts to invariance of the trace under adjoint action.    
One can also introduce the inner product
\be
(F_1,F_2) = \int F_1\dag F_2.
\ee

A complete basis of functions on $S_N^2$ is given by
the $(N+1)^2$ spherical harmonics, 
$Y^J_j, (J = 0, 1, ..., N;  -J \leq j \leq J)$ \footnote{We will
use capital and small letter (e.g. $(J,j)$) to refer to the 
eigenvalue of the angular momentum operator $\bf{J^2}$ and $J_z$
respectively.}, which are the weight basis of the spin $J$ component
of $S_N^2$ explained above.
They correspond to the usual spherical harmonics, however 
the angular momentum has an upper bound $N$ here. This is a
characteristic feature of fuzzy sphere. 
The normalization and reality for these matrices can be taken to be
\be
(Y^J_j, Y^{J'}_{j'}) = \delta_{JJ'} \delta_{jj'}, \qquad
(Y^J_j)^\dagger = (-1)^J Y^J_{-j}.
\ee
They obey the ``fusion'' algebra
\be \label{YY}
Y^I_i Y^J_j = 
\sqrt{\frac{N+1}{4\pi}}
\sum_{K,k} (-1)^{2\a+I+J+K+k} \sqrt{(2I+1)(2J+1)(2K+1)}
\left( \begin{array}{ccc} I&J&K\\ i&j&-k \end{array} \right)          
\left\{\begin{array}{ccc} I&J&K\\ \a&\a&\a \end{array} \right\} \; Y^K_k ,
\ee
where the sum is over $0 \leq K \leq N, -K \leq k \leq K$, and 
\be
\a = N/2 . 
\ee
Here the first bracket is the  Wigner $3j$-symbol
and the curly bracket is the $6j$--symbol of
$su(2)$, in the standard mathematical normalization
\cite{Var}. Using the Biedenharn--Elliott identity \eq{BE}, 
it is easy to show that \eq{YY} is associative.
In particular, 
$Y^0_0 = \frac{1}{\sqrt{4\pi}}\; \bf{1}$.
The relation \eq{YY} is independent of the radius $R$, but depends on
the deformation parameter $N$. It is  
a deformation of the algebra of product of the spherical harmonics on the
usual sphere.  We will need \eq{YY} to derive the form of
the propagator and vertices in the angular momentum basis. 

Now we turn to various limits of the fuzzy sphere. By tuning the
parameters $R$ and $N$, one can obtain different limiting algebras of
functions. In particular, we consider the commutative sphere $S^2$ and 
the noncommutative plane $\bR^2_\th$.

\subsection{The commutative sphere limit $S^2$}

The commutative limit is defined by 
\be \label{limit-s}
N \rightarrow \infty ,\quad \mbox{keeping $R$ fixed}.
\ee
In this limit, \eq{def1} reduces to $[x_i,x_j]
=0$ and
we obtain the commutative algebra of functions on the usual sphere $S^2$. 
Note that \eq{YY} reduces to the standard product of spherical
harmonics, due to the asymptotic relation between the $6j$--symbol and 
the Wigner $3j$-symbol \cite{Var}, 
\be
\lim_{\a\rightarrow \infty} (-1)^{2\a} \sqrt{2 \a}
\left\{\begin{array}{ccc} I&J&K\\ \a&\a&\a \end{array} \right\} 
= (-1)^{I+J+K}
\left( \begin{array}{ccc} I&J&K\\ 0&0&0 \end{array} \right) .
\ee
 
\subsection{The quantum plane limit $\bf{R}^2_\th$ }

If the fuzzy sphere is blown up around a given point, it 
becomes an approximation of the quantum plane \cite{madore}. 
To obtain this planar limit, it is convenient to
first introduce an alternative representation of the fuzzy sphere
in terms of stereographic projection. Consider the generators
\be
y_+ = 2 R x_+ (R-x_3)^{-1}, \quad y_- = 2 R (R-x_3)^{-1} x_-,
\ee
where $x_\pm = x_1 \pm i x_2$. The generators 
$y_\pm$  are the coordinates of the stereographic projection 
from the north pole. $y=0$ corresponds to the south pole. Now we take
the large $N$ and large $R$ limit, such that
\be \label{limit-p}
N \to \infty, \quad R^2 = N \th /2 \to \infty, 
\quad \mbox{keeping $\th$ fixed}. 
\ee
In this limit, 
\be
\frac{\l_N}{\sqrt{\th}}\; \sim \;  \frac 1{\sqrt{N}}
\ee
and $[y_+,y_-] = -4 R^2 \l_N (R-x_3)^{-1} + o(\l_N^2)$.
Since $y_+ y_- = 4R^2 (R+x_3) (R-x_3)^{-1} + o(N^{-1/2})$, 
we can cover the whole $y$-plane with
$x_3 = -R + \b/R$ with finite but arbitrary $\b$.
The commutation relation of the $y$ generators takes the form
\be
[y_+,y_-] = -2 \th
\ee
up to corrections of order $\l_N^2$, or
\be
[y_1, y_2] = -i \th
\ee
with $y_\pm = y_1 \pm i y_2$. 

\section{One Loop Dynamics of $\Phi^4$ on the Fuzzy Sphere}

Consider a scalar $\Phi^4$ theory on the fuzzy sphere, with action
\be
S_0 = \int \frac 12\; \Phi (\Delta + \m^2) \Phi + 
    \frac{g}{4 !} \Phi^4.
\ee
Here $\Phi$ is Hermitian, $\m^2$ is the dimensionless  
mass square, $g$ is a dimensionless coupling
and $\Delta = \sum J_i^2$ is the Laplace operator.
The differential operator $J_i$ acts on function $F \in S_N^2$
as
\be
J_i F = \frac{1}{\l_N}[x_i, F].
\ee
This action is valid for any radius $R$, since $\m$ and
$g$ are dimensionless. To quantize the theory, 
we will follow the path integral quantization procedure as explained in
\cite{gkp}. We expand $\Phi$ in terms of the modes, 
\be
\Phi = \sum_{L,l} a^L_{l} Y^L_l, \quad 
a^{L}_l{}\dag = (-1)^l a^L_{-l}.
\ee
The Fourier
coefficient $a^L_l$ are then treated as the dynamical variables, and 
the path integral
quantization is defined by integrating over all possible configuration of
$a^L_l$. Correlation functions are computed using \cite{gkp}
\be
\langle a^{L_1}_{l_1}   \cdots  a^{L_k}_{l_k} \rangle = 
\frac{\int [\cD \Phi] e^{-S_0}\; a^{L_1}_{l_1}   \cdots  a^{L_k}_{l_k}  }
{\int [\cD \Phi] e^{-S_0}}.
\ee
For example, the propagator is
\be \label{prop}
\langle a^{L}_{l}   a^{L'}_{l'}{}\dag \rangle =
(-1)^l\langle a^{L}_{l}   a^{L'}_{-l'} \rangle =
\d_{L L'} \d_{l l'} \frac{1}{L(L+1) + \mu^2 },
\ee
and the vertices for the $\Phi^4$ theory are given by
\be
a^{L_1}_{l_1} \cdots  a^{L_4}_{l_4}\;  V(L_1,l_1; \cdots; L_4,l_4)
\ee
where
\bea \label{V}
 V(L_1,l_1; \cdots; L_4,l_4)&& = \frac{g}{4!} \frac{N+1}{4 \pi} 
 (-1)^{L_1+L_2+L_3+L_4} \prod_{i=1}^4 (2L_i+1)^{1/2}  \sum_{L,l}
 (-1)^{l}(2L+1)  \nn\\ 
&& \cdot\left( \begin{array}{ccc} L_1&L_2&L\\ l_1&l_2&l \end{array} \right)
\left( \begin{array}{ccc} L_3&L_4&L\\ l_3&l_4&-l \end{array} \right) 
\left\{\begin{array}{ccc} L_1&L_2&L\\ \a&\a&\a \end{array} \right\} 
\left\{\begin{array}{ccc} L_3&L_4&L\\ \a&\a&\a \end{array} \right\} .
\eea
One can show that $V$ is symmetric with respect to cyclic
permutation of its arguments $(L_i,l_i)$.

The  $1PI$ two point function at one loop is obtained by
contracting 2 legs in \eq{V} using the propagator \eq{prop}.
The planar contribution is defined by contracting neighboring legs:
\be \label{aap}
(\Gamma^{(2)}_{planar})^{L L'}_{l l'}
= \frac{g}{4\pi} \frac 13\; \d_{L L'} \d_{l, -l'}(-1)^{l} 
\cdot I^P, 
\quad
I^P:= \sum_{J=0}^N \frac{2J+1}{J(J+1) +\m^2} .
\ee
All 8 contributions are identical.
Similarly by contracting non--neighboring legs, we find 
the non--planar contribution
\be \label{aanp}
(\Gamma^{(2)}_{nonplanar})^{L L'}_{l l'}
=  \frac{g}{4 \pi} \frac 16\; \d_{L L'} \d_{l, -l'} (-1)^{l} \cdot I^{NP},
\quad
I^{NP}:= \sum_{J=0}^N (-1)^{L+J+2\a} \;\frac{(2J+1)(2\a+1)}{J(J+1) +\m^2 } 
\left\{\begin{array}{ccc} \a&\a&L\\ \a&\a&J \end{array} \right\}.
\ee
Again the 4 possible contractions agree.
These results can be found using standard identities for the 
$3j$ and $6j$ symbols, see e.g. \cite{Var} and Appendix B.

It is instructive to note that $I^{NP}$ 
can be written in the form  
\be
I^{NP}= \sum_{J=0}^N \frac{2J+1}{J(J+1) +\m^2} \; f_J ,
\ee
where $f_J$ is obtained from the generating function
\be
f(x) = \sum_{J=0}^\infty f_J x^J
=\frac{1}{1-x} \;
{}_2F_1 (-L,L+1,2 \a+2, \frac{x}{x-1}) 
{}_2F_1 (-L,L+1, -2 \a, \frac{x}{x-1}).
\ee
Here the hypergeometric function ${}_2F_1(-L,L+1; c;z)$ is a
polynomial of degree $L$ for any $c$. 
 Note that the
oscillatory sign in $I^{NP}$ in \eq{aanp} is cancelled by the sign 
of the $6j$-symbol in \eq{racah_app}, and is replaced by a  slower 
oscillatory behaviour of the  $6j$-symbol as a function of $L$ and
$J$. The latter is precisely the counter-part of
the nonplanar Moyal phase factor in the noncommutative plane 
case
\footnote{
It was argued  in \cite{vaidya} that the nonplanar
two-point function has a different sign for even and odd 
external angular momentum $L$. 
This is not correct, because
as we just explained, the oscillations are indeed much milder
after combining with the $6j$-symbol.  As we will show, this
leads to a well--behaved loop integral which is a sum over all $J$, 
and the resulting low $L$--behaviour is completely regular. 
We note that only the single value of $J=2\a$ 
was considered in \cite{vaidya}.
} .
For example, for $L=0$, one obtains 
\be
f_J =1, \quad \mbox{$0 \leq J \leq N$},
\ee 
and hence the planar and nonplanar two point functions coincides.
For $L=1$, we have
\be
f_J = 1-\frac{J(J+1)}{2 \a (\a+1)}, \quad \mbox{$0 \leq J \leq N$ },
\ee
and hence 
\be \label{L1}
I^{NP} = I^{P} - \frac{1}{2 \a(\a+1)}
\sum_{J=0}^{2\a} 
\frac{J (J+1) (2J+1)}{J(J+1) +\m^2} .
\ee
Note that 
the difference between the planar and nonplanar two point functions is 
finite. It is easy to convince oneself that for any finite external
angular momentum $L$, the difference between the planar and nonplanar
two point function is finite and analytic in $1/ \a$. 
This fact is important as it implies that, unlike in the $\bR^n_\th$ case, 
there is no infrared singularity developed in the nonplanar 
amplitude. We will have more to say about this later.

\subsection{On UV/IR mixing and the commutative limit}

Let us recall that in the case of noncommutative space $\bR^{n}_\th$,
the one--loop contribution to the effective action often
develops a singularity at $\th p =0$ \cite{uvir1,vac}. 
This infrared singularity is generated by
integrating out the infinite number of 
degrees of freedom in the nonplanar loop. 
This phenomenon is referred to as ``UV/IR mixing'', and it implies 
in particular that  
(1)  the nonplanar amplitude is singular when the external 
momentum is zero in the noncommutative directions;
and 
(2) the quantum effective action in the commutative limit is different
from the quantum effective action of the commutative limit \cite{wzw}.

\underline{Effective action on the fuzzy sphere}

We want to understand the behavior of the corresponding planar and 
nonplanar two point functions on the fuzzy sphere, to see if there is 
a similar UV/IR phenomenon. We emphasize that this is not obvious a priori 
even though quantum field theory on the fuzzy sphere is always finite.
The question is whether the 2--point function is smooth at 
small values of $L$, or rapidly oscillating as was indeed claimed in 
a recent paper \cite{vaidya}. 
Integrating out all the degrees of freedom in the
loop could in principle generate a IR singularity, for large $N$.

However, this is not the case.  
We found above  that the planar
and nonplanar two point function agree precisely with each 
other when the external angular momentum $L=0$. 
For general $L$, a closed expression for $f_J$ 
for general $L$ is difficult to obtain. We will derive below an approximate 
formula for the difference $I^{NP} - I^{P}$, which is found to 
be an excellent approximation for large $N$ by numerical tests, and 
becomes exact in the  commutative limit $N \rightarrow \infty$.

First, the planar contribution to the two point function 
\be
I^P= \sum_{J=0}^N \frac{2J+1}{J(J+1) +\m^2} 
\ee
agrees precisely with the corresponding terms on the classical
sphere as $N \rightarrow \infty$, and it diverges
logarithmically
\be
I^P  \sim \log \a + o(1).
\ee
To understand the nonplanar contribution, 
we start with the following approximation formula \cite{Var} 
for the $6j$ symbols due to Racah,
\be
\left\{\begin{array}{ccc} \a&\a&L\\ \a&\a&J \end{array} \right\} \approx
\frac{(-1)^{L+ 2 \a +J}}{2\a} P_L (1-\frac{J^2}{2 \a^2}),
\label{racah_app}
\ee
where $P_L$ are the Legendre Polynomials.
This turns out to be an excellent approximation for 
all $0 \leq J \leq 2\a$, provided
$\a$ is large and $L \ll \a$. 
Since this range of validity of this 
approximation formula is crucial for us, we shall derive it in Appendix A.
This allows then to rewrite the sum in \eq{aanp} to a very good
approximation as
\be
I^{NP} - I^{P} = \sum\limits_{J = 0}^{2\a} \frac{2J+1}{J(J+1) + \m^2}
\left(P_L(1-\frac{J^2}{2 \a^2}) -1 \right)
\ee
for large $\a$. Since $P_L(1) = 1$ for all $L$, only $J \gg 1$
contributes, and one can approximate the sum by the integral
\bea
I^{NP} - I^{P} &&\approx \int\limits_0^{2} du \;
          \frac{2u + \frac{1}{\a}}{u^2 + \frac{u}{\a}+ \frac{\m^2}{\a^2}}
  \left(P_L(1-\frac{u^2}{2}) -1 \right) \nn\\
 &&= \int\limits_{-1}^{1}dt \; \frac{1}{1-t} (P_L(t) - 1) + o (1/\a),
\eea
assuming $\m \ll \a$.
This integral is finite for all $L$. 
Indeed using generating functions techniques,
it is easy to show that
\be
\int\limits_{-1}^{1}dt  \frac{1}{1-t} (P_L(t) - 1) = 
  -2\; (\sum_{k=1}^L \frac 1k) = -2 h(L),
\label{Legendre-int}
\ee 
where $h(L) = \sum_{k=1}^L \frac 1k$ is the harmonic number and $h(0) =0$. 
While $h(L) \approx log L$ for large $L$, it is finite and 
well--behaved for small $L$. Therefore
we obtain the effective action 
\be
 S_{one-loop} = S_0 + 
           \int \frac 12\;  \Phi (\delta \mu^2 - 
          \frac{g}{12\pi} h(\widetilde\Delta)) \Phi + o(1/\a)  
\ee
to the first order in the coupling where
\be 
\delta \mu^2 = \frac{g}{8\pi} \sum_{J=0}^N \frac{2J+1}{J(J+1) +\m^2}
\ee 
is the mass square renormalization, and $\widetilde\Delta$ 
is the function of the Laplacian which has eigenvalues
$L$ on $Y^L_l$. Thus we find that the effects due
to noncommutativity are analytic in the noncommutative parameter
$1/\a$.
This is a finite quantum effect with nontrivial, but mild $L$ dependence.
Therefore no IR singularity is
developed, and we conclude that there is no UV/IR problem on the
fuzzy sphere
\footnote{
The author of \cite{vaidya}  
adopted a Wislonian approach integrating the "cutoff" by one unit, and
argued that the effective
action is not a smooth function of the external momentum and suggested
this to be a signature of UV/IR mixing.  We disagree with his result.
In this paper, we follow the more conventional program 
of renormalization (for the 2-point function),
and  calculate the full loop
integral  which is perfectly  regular. 
}.

\underline{The commutative limit}

The commutative limit of the QFT is defined by the limit 
\be
\a \to \infty, \quad\mbox{keeping $R$, $g$, $\m$ fixed}.
\ee
In this limit, the resulting
one-loop effective action differs from the effective action obtained
by quantization on the commutative sphere by an amount 
\be  \label{NCA}
\Gamma^{(2)}_{NCA} =  - \frac{g}{24\pi} \int \Phi h(\widetilde\Delta) \Phi.
\ee
We refer to this as a ``NonCommutative Anomaly'', 
since it is the piece of the quantum effective action which 
is slightly nonlocal and therefore not present in the classical action. 
``Noncommutative'' also
refers to fact that the quantum effective action depends on 
whether we quantize first or take the commutative limit first.

The new term $\Gamma^{(2)}_{NCA}$ modifies the dispersion relation on the 
fuzzy sphere.
It is very remarkable that such a ``signature'' of 
an underlying noncommutative space exists, even as the noncommutativity
on the geometrical level is sent to zero. A similar phenomena is the
induced Chern-Simon term in 3-dimensional gauge theory on $\bR^3_\th$
\cite{wzw}. 
This has important 
implications on the detectability of an underlying noncommutative structure.
The reason is that the vacuum fluctuations ``probe'' the structure
of the space even in the UV, 
and depend nontrivially on the external momentum  
in the nonplanar diagrams. 
Higher--order corrections may modify the result.
However since the theory is completely
well--defined for finite $N$, the above result  
\eq{NCA} is meaningful for small
coupling $g$.

Summarizing, we find  
that quantization and taking the commutative limit does not commute 
on the fuzzy sphere, a fact which we refer to as 
``noncommutative anomaly''. A similar phenomenon also occurs on the
noncommutative quantum plane $\bR^n_\th$. 
However, in contrast to the case of the quantum plane, the
``noncommutative anomaly'' here
is not due to UV/IR mixing since there is
no UV/IR mixing on the fuzzy sphere.
We therefore suggest that the existence of a ``noncommutative anomaly'' 
is a generic phenomenon and is independent of UV/IR 
mixing\footnote{However, 
as we will see, they are closely related.
We will show in the next section that the UV/IR mixing on the 
noncommutative plane arises in the planar limit of the ``NCA'' for the fuzzy
sphere
}. 
One can expect that the ``noncommutative anomaly'' does not occur for
supersymmetric theories on the 2--sphere. 

\section{Planar Limit of Quantum $\Phi^4$}

In this section, we study the planar limit 
of the $\Phi^4$ theory on the fuzzy sphere at one loop. 
Since we
have shown that there is no UV/IR mixing on the fuzzy sphere, one may
wonder whether \eq{sphere-reg} could provide a regularization for the
nonplanar two point function  \eq{planeI2} on $\bR^2_\th$ which does
not display an infrared singularity. This would be very nice,
as this would mean that UV/IR can be understood as an artifact that arises
out of a bad choice of variables. However, this is not the case.

To take the planar limit, we need in addition
to  \eq{limit-p}, also 
\be
\mu^2 = m^2 R^2 \sim \a \to \infty, \quad 
     \mbox{keeping  $m$ fixed},
\ee
so that a massive scalar theory is obtained. 
We wish to identify in the limit of large $R$ the 
modes on the sphere with angular
momentum $L$ with modes on the plane with linear momentum $p$. This can
be achieved by matching the Laplacian on the plane 
with that on the sphere in the large radius limit, ie.
\be \label{Lp}
L (L+1)/R^2 = p^2.
\ee
It follows that
\be \label{pR}
p = \frac{L}{R}. 
\ee
Note that by (\ref{limit-p}), 
a mode with a fixed nonzero $p$ corresponds to a mode on the
sphere with large $L$:
\be \label{Lfinitep}
L \sim R \sim \sqrt{\a}. 
\ee
Since $L$ is bounded by $\a$, there is a UV cutoff $\L$ on the plane at
\be
\L = \frac{2\a}{R}.
\ee
Denote the external momentum of the two point function by $p$. 
It then follows that $\a \gg L \gg 1$ as long as $p \neq 0$. 

It is easy to see that the planar amplitude \eq{aap} becomes 
\be
I^P = 2 \int_0^\L dk \frac{k}{k^2+m^2} 
\ee
in the quantum plane limit, with $k = J/R$. This is
precisely the planar contribution to the two point function on $\bR^2_\th$. 

For the nonplanar two point function \eq{aanp}, 
we can again use the formula (\ref{racah_app}) which is valid for all
$J$ and large $\a$, since the condition $\a \gg L$ is guaranteed by 
(\ref{Lfinitep}). 
Therefore 
\be \label{I-int-racah}
I^{NP}(p) = 2 \int_{0}^\L dk\; \frac{k}{k^2+m^2} 
       P_{p R}(1-2\frac{k^2}{\L^2})
\ee
For large $L = p R$, we can use the approximation formula
\be
P_L(\cos \phi) = \sqrt{\frac{\phi}{\sin \phi}}J_0((L+1/2)\phi) \quad
  +\; O(L^{-3/2}),
\label{PLJ-approx}
\ee
which is uniformly convergent \cite{magnus} 
as $L \rightarrow \infty$ in the interval 
$0 \leq \phi \leq \pi - \epsilon$ for any small, but finite $\epsilon>0$.
Then one obtains
\bea 
I^{NP}(p) &\approx&  2\int_{0}^\L dk \;\frac{k}{k^2+m^2} 
                   \sqrt{\frac{\phi_k}{\sin \phi_k}}\; J_0(p R \phi_k) \nn\\
          &\approx&  2\int_{0}^\L dk \;\frac{k}{k^2+m^2}\; J_0(\th p k), 
\label{sphere-reg}
\eea
where $\phi_k = 2 \arcsin(k/\L)$. The singularity at $\phi = \pi$ on the rhs 
of \eq{PLJ-approx} (which is an artefact of the approximation and not 
present in the lhs) is integrable and does not contribute to \eq{sphere-reg}
for large $p\L\th$. 
The integrals in \eq{sphere-reg} are (conditionally) 
convergent for $p\neq 0$, and the approximations become exact
for $p\L\th \rightarrow \infty$. Therefore we recover
precisely the same form as the one loop nonplanar two point function on
$\bR^2_\th$,
\be \label{planeI2}
\frac{1}{2 \pi} \int_{0}^\L d^2 k \; 
      \frac{1}{k^2+m^2} e^{i \th p \times k}.
\ee
For small $p\L\th$, i.e. in the vicinity of the induced infrared divergence
on $\bR^2_\th$,
these approximations are less reliable. We can obtain
the exact form of the infrared divergence from \eq{Legendre-int},
\be
I^{NP} = - 2 \log(p\sqrt{\th}) + (I^P - \log{\a}).
\ee
Hence we find the same logarithmic singularity in the infrared
as on $\bR^2_\th$ \cite{uvir1}. In
other words, we find that the UV/IR mixing phenomenon
which occurs in QFT on $\bR^2_\th$ can be understood as the infinite 
limit of the noncomutative anomaly \eq{NCA} on the fuzzy sphere.
Hence one could use the fuzzy sphere as a regularization
of $\bR^2_\th$, where the logarithmic singularity 
$\log(p\sqrt{\th})$ gets ``regularized'' by \eq{Legendre-int}.

\section{Discussion}

We have done a careful analysis of the one--loop dynamics 
of scalar $\Phi^4$ theory on the fuzzy sphere $S^2_N$. We found that the 
two point function is completely regular, without any UV/IR
mixing. We also give a closed expression for 
the two point function in the commutative limit, i.e. we find an exact form
for the leading term in a $1/N$ expansion.
Using this we discover a ``noncommutative anomaly'' 
(NCA), which characterizes the difference between  
the quantum effective action on the commutative sphere $S^2$ and
the commutative limit $N \rightarrow \infty$
of the quantum effective action on the fuzzy sphere. 
This anomaly is finite but mildly nonlocal on $S^2$, and changes the 
dispersion relation. 
It arises from the nonplanar loop integration. 

It is certainly intriguing and perhaps disturbing
that even an ``infinitesimal'' quantum structure of
(space)time has a finite, nonvanishing 
effect on the quantum theory. Of course this
was already found in the UV/IR phenomenon on $\bR^n_\th$, 
however in that case
one might question whether the quantization procedure based on deformation
quantization is appropriate. On the fuzzy sphere, the
result is completely well-defined and unambiguous.
One might argue that a ``reasonable'' QFT should be free of such a
NCA, so that  the effective, macroscopic theory is insensitive to
small variations of the structure of spacetime at short distances. 
On the other hand, it is conceivable that 
our world is actually noncommutative, and the noncommutative dynamics 
should be taken seriously. Then  there is no reason to
exclude theories with NCA. In particular, one would like to 
understand better how sensitive these ``noncommutative anolamies''
are to the detailed quantum structure of spacetime.

By approximating the QFT on $\bR^2_\th$ with the QFT on the fuzzy
sphere, we can explain the UV/IR mixing from the point of view of the
fuzzy sphere as a infinite variant of the NCA. 
In some sense, we have regularized $\bR^2_\th$.
It would be interesting
to provide an explanation of the UV/IR mixing also for the higher
dimensional case $\bR^4_\th$. To do this, the first step is to realize
$\bR^4_\th$ as a limit of a ``nicer'' noncommutative
manifold. A first candidate is the product of two fuzzy spheres. Much
work remains to be done to clarify this situation. 

It would also be very desirable to include fermions and gauge fields in 
these considerations. In particular it will be interesting to determine 
the dispersion relation for ``photons'', depending on the 
``fuzzyness'' of the underlying geometry.
In the case of noncommutative QED on $\bR^4_\th$, this question was studied 
in \cite{vac},
where a nontrivial modification to the dispersion relation of the ``photon''
was found which makes the theory ill--defined.
In view of our results, 
one may hope that these modifications are milder on the fuzzy sphere 
and remain physically sensible.

\section*{Acknowledgements} 

We would like to thank H. Grosse and R. Russo for stimulating discussions.
H. S. also thanks the DFG for a fellowship, as well as D. Schiff for 
hospitality at the Laboratoire de Physique Th\'eorique in Orsay.

\section*{Appendix A}

We derive the approximation formula (\ref{racah_app}) for large $\a$
and $0 \leq J \leq 2\a$, assuming $L \ll \a$. 

There is an exact formula for the $6j$ coefficients due to Racah
(see e.g. \cite{Var}), which can be written in the form
\be 
\left\{\begin{array}{ccc} \a&\a&L\\ \a&\a&J \end{array} \right\} =
(-1)^{2\a+J}\sum\limits_{n} (-1)^n 
           \left(\begin{array}{c}L \\ n\end{array}\right)^2
   \frac{(2\a-L)!(2\a+J+n+1)!(2\a-J)!(J!)^2}
   {(2\a+L+1)!(2\a+J+1)!(2\a-J-n)!((J-L+n)!)^2}.
\ee
The sum is from $n = \max\{0,L-J\}$ to $\min\{L,2\a-J\}$,
so that all factorials are non-negative.
Assume first that $L \leq J \leq 2\a-L$, so that the 
sum is from $0$ to $L$. 
Since $\a \gg L$, this becomes
\be
\left\{\begin{array}{ccc} \a&\a&L\\ \a&\a&J \end{array} \right\} \approx
(-1)^{2\a+J} \frac 1{(2\a)^{2L+1}} \sum\limits_{n=0}^L (-1)^n 
  \left(\begin{array}{c}L \\ n\end{array}\right)^2
  (4\a^2 - J^2)^n \left(\frac{J!}{(J-(L-n))!}\right)^2,
\label{6j-deriv}
\ee
dropping corrections of order $o(\frac L{\a})$.
Now there are 2 cases: either $J \gg L$, or otherwise $J \ll \a$
since $\a \gg L$. Consider first
\begin{enumerate}
\item  $J \gg L$:

Then $\frac{J!}{(J-(L-n))!}$ can be replaced by $J^{L-n}$, up to 
corrections of order $o(\frac LJ)$. Therefore 
\bea
\left\{\begin{array}{ccc} \a&\a&L\\ \a&\a&J \end{array} \right\} &\approx&
  \frac{(-1)^{2\a+J}}{2\a} \left(\frac J{2\a}\right)^{2L}
  \sum\limits_{n=0}^L (-1)^n  \left(\begin{array}{c}L \\ n\end{array}\right)^2
  \left( (\frac{2\a}J)^2 - 1 \right)^n \nn\\
  &=& \frac{(-1)^{2\a+J}}{2\a} P_L(1- \frac{J^2}{2\a^2}),
\eea
as claimed.
\item $J \ll \a$: 

Then in the sum (\ref{6j-deriv}), 
the dominant term is $n = L$, because 
$\frac{J!}{(J-(L-n))!} \leq J^{L-n}$. Therefore one can safely
replace the term $\frac{J!}{(J-(L-n))!}$ in this sum by its value at $n=L$,
namely $J^{L-n}$. The remaining terms are smaller by a 
factor of $(\frac J{\a})^2$. Hence we can continue as in case 1.

\end{enumerate}

If  $J \leq L$, one 
can either use the same argument as in the 2nd case since the
term $n = L$ is dominant, or use the symmetry of the 
$6j$ symbols in $L,J$ together with 
$P_L(1- \frac{J^2}{2\a^2}) \approx  P_J(1- \frac{L^2}{2\a^2})$
for $J,L \ll \a$.
Finally if $J+L \geq 2\a$, then the term $n=0$ dominates, and 
one can proceed as in case 1. Therefore
(\ref{racah_app}) is valid for all $0 \leq J \leq 2\a$.

One can illustrate
the excellent approximation for the $6j$ symbols provided 
by (\ref{racah_app}) for all $0 \leq J \leq 2\a$ using numerical calculations.

\section*{Appendix B} 

We quote here some identities of the $3j$ and $6j$ symbols which are 
used to derive the expressions (\ref{aap}) and (\ref{aanp})
for the one--loop corrections.
The $3j$ symbols satisfy the orthogonality relation
\be
\sum_{j,l}
\left( \begin{array}{ccc} J&L&K\\ j&l&k \end{array} \right)
\left( \begin{array}{ccc} J&L&K'\\ -j&-l&-k' \end{array} \right)
= \frac{(-1)^{K-L-J}}{2K+1} \d_{K, K'} \d_{k, k'},
\ee
assuming that $(J,L,K)$ form a triangle. 

The $6j$ symbols satisfy standard symmetry properties, and
the orthogonality relation
\be
\sum_{N} (2N+1)
    \left\{ \begin{array}{ccc} A&B&N\\ C&D&P \end{array} \right\}
    \left\{ \begin{array}{ccc} A&B&N\\ C&D&Q  \end{array} \right\}
  = \frac 1{2P+1}\; \d_{P, Q} ,
\ee
assuming that $(A,D,P)$ and $(B,C,P)$ form a triangle.
Furthermore, the following sum rule is used in \eq{aanp}
\be
\sum_{N} (-1)^{N+P+Q} (2N+1) 
  \left\{ \begin{array}{ccc} A&B&N\\ C&D&P  \end{array} \right\}
  \left\{ \begin{array}{ccc} A&B&N\\ D&C&Q  \end{array} \right\}
 = \left\{ \begin{array}{ccc} A&C&Q\\ B&D&P  \end{array} \right\}.
\ee
The Biedenharn--Elliott relations are needed to verify associativity of
(\ref{YY}):
\be \label{BE}
\sum_{N} (-1)^{N+S} (2N+1) 
 \left\{ \begin{array}{ccc} A&B&N\\ C&D&P  \end{array} \right\}
 \left\{ \begin{array}{ccc} C&D&N\\ E&F&Q  \end{array} \right\}
 \left\{ \begin{array}{ccc} E&F&N\\ B&A&R  \end{array} \right\} = 
 \left\{ \begin{array}{ccc} P&Q&R\\ E&A&D  \end{array} \right\}
 \left\{ \begin{array}{ccc} P&Q&R\\ F&B&C  \end{array} \right\},
\ee
where $S = A+B+C+D+E+F+P+Q+R$.
All these can be found e.g. in \cite{Var}.

\ed